\pdfoutput=1

\documentclass[11pt]{article}

\usepackage[final]{acl}

\usepackage{times}
\usepackage{latexsym}

\usepackage[T1]{fontenc}

\usepackage[utf8]{inputenc}

\usepackage{microtype}

\usepackage{inconsolata}

\usepackage{graphicx}

\usepackage{multirow}
\usepackage{booktabs}
\usepackage{amssymb}
\usepackage{subfigure}
\usepackage{color}

\newcommand{\our}{\textsc{SEAL}}
\newcommand{\dataset}{StructDocRetrieval}

\title{
SEAL: Structure and Element Aware Learning to Improve Long Structured Document Retrieval
}

\author{
Xinhao Huang$^{1}$\thanks{~~Equal Contribution},
Zhibo Ren$^{3}$\textsuperscript{*},
Yipeng Yu$^{3}$,
Ying Zhou$^{4}$,
Zulong Chen$^{3}$\thanks{~~Corresponding Author},
Zeyi Wen$^{1,2}$\textsuperscript{\dag} \\
\textsuperscript{\rm1}HKUST (Guangzhou), Guangzhou, China \quad
\textsuperscript{\rm2}HKUST, Hong Kong, China~~~\\
\textsuperscript{\rm3}Alibaba Group, Hangzhou, China~~~ \quad
\textsuperscript{\rm4}Zhejiang Lab, Hangzhou, China~~~
\\
\texttt{chenzulong198867@gmail.com, wenzeyi@ust.hk}
}

\begin{document}
\maketitle
\begin{abstract}

In long structured document retrieval, existing methods typically fine-tune pre-trained language models (PLMs) using contrastive learning on datasets lacking explicit structural information. This practice suffers from two critical issues: 1) current methods fail to leverage structural features and element-level semantics effectively, and 2) the lack of datasets containing structural metadata.
To bridge these gaps, we propose \our, a novel contrastive learning framework. It leverages structure-aware learning to preserve semantic hierarchies and masked element alignment for fine-grained semantic discrimination. Furthermore, we release \dataset, a long structured document retrieval dataset with rich structural annotations. Extensive experiments on both released and industrial datasets across various modern PLMs, along with online A/B testing, demonstrate consistent performance improvements, boosting NDCG@10 from 73.96\% to 77.84\% on BGE-M3.
The resources are available at this \href{https://github.com/xinhaoH/SEAL}{URL}.

\end{abstract}

\section{Introduction}
 
\begin{figure}[t]
\centering
\includegraphics[width=0.99\columnwidth]{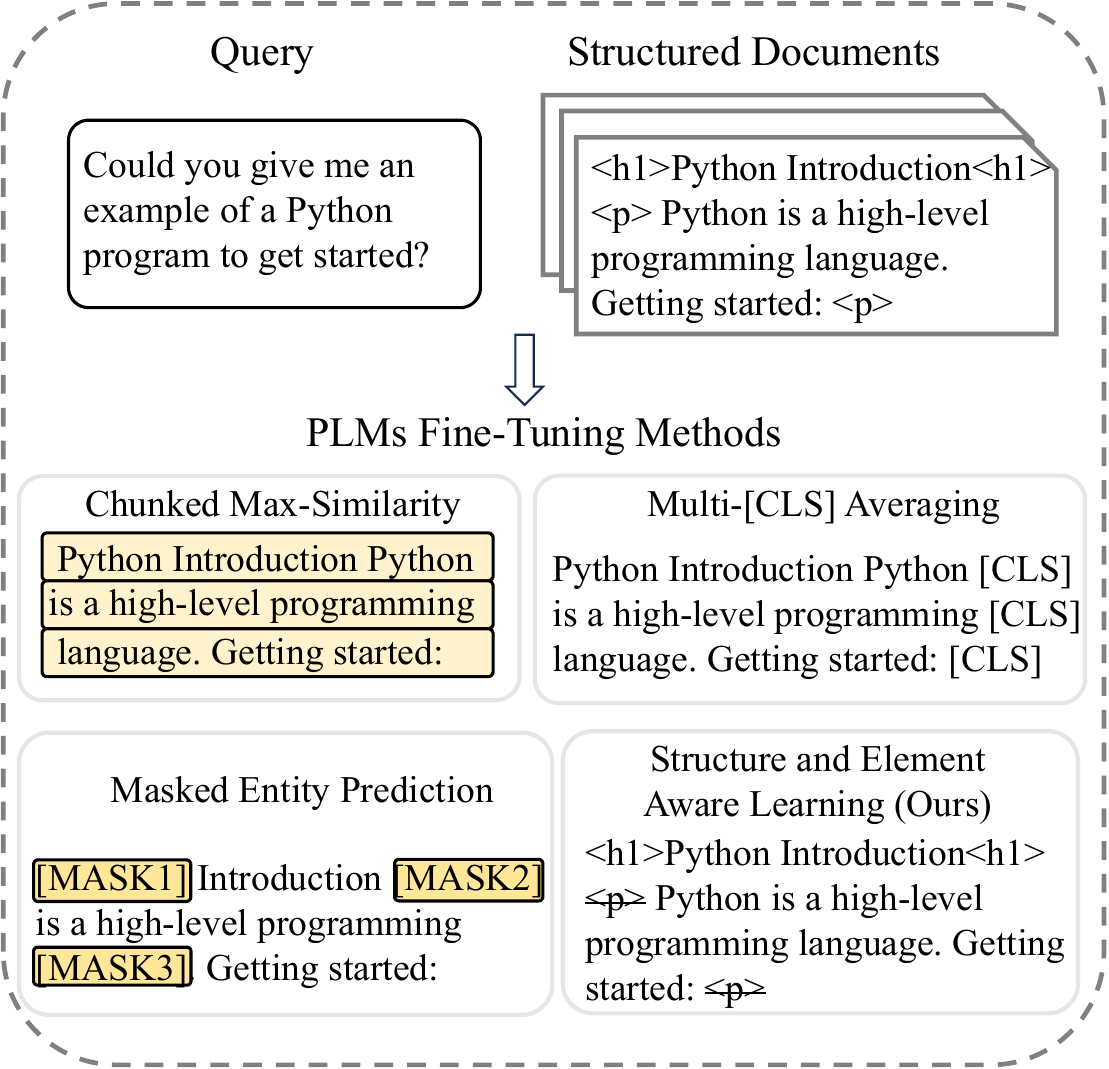} 
\caption{An overview of long structured documents retrieval methods.}
\label{fig:intro}
\end{figure}

Document retrieval is a fundamental component of knowledge-intensive systems, such as Retrieval-Augmented Generation (RAG) \cite{survey, RAG_Survey}. Despite recent advances in PLMs that extend sequence processing capacity (e.g., from 512 to 8192 tokens), the precise identification of query-relevant content in long documents remains an open challenge \cite{BERT, M3_Embedding}. Existing methods typically employ contrastive learning trained on query and raw textual content \cite{MDR, ANCE, DANCE, SANTA, CodeT5, CodeRetriever, rao2022reproducibility, rao2023dynamic, apt} or to optimize PLM representations. Nevertheless, as illustrated in Figure \ref{fig:intro}, this paradigm exhibits two key limitations: (1) Structural blindness arising from raw text processing that disrupts document hierarchies and discards semantic markup indicators like H1/H2 headings \cite{html}; and (2) Insufficient element-level alignment capacity over fragmented text segments, which fails to preserve fine-grained semantic relationships. Furthermore, the lack of structural metadata in current datasets leaves long structured document retrieval scenarios under-studied. 

To address these limitations, we propose \our, a novel contrastive learning framework that integrates structural semantics through two key components: (1) A structure-aware contrastive learning method leveraging HTML transformation and structural tag inclusion/exclusion enables semantic hierarchy induction; and (2) An element-level alignment mechanism employing stochastic element masking forces the model to achieve granular semantic alignment. Our SEAL assigns significantly higher relevance to important elements (e.g., chapter titles or intentionally bolded query terms) compared to mentions in body text. This contrasts with structure-agnostic retrieval, which erroneously assigns them equal weights.

Beyond the methodological challenges, the availability of suitable benchmark datasets is crucial. Current retrieval datasets either focus on task-specific retrieval (e.g., passage, product, code, ranking) \cite{PassageRetrieval, ProductSearch, CodeSearch} or lack structured metadata \cite{MSMARCO}. Their short text lengths (typically <1,000 words) further render their usage.
To bridge this gap and provide a reproducible resource for the community, we release \dataset, a dataset designed for long structured document retrieval. \dataset{} contains annotated documents with explicit structural semantics and an average length of more than 10,000 words. 

We conduct extensive experiments to evaluate \our{} against the state-of-the-art document retrieval methods on \dataset{} and the industrial dataset, using different modern PLMs. We also validate its practical effectiveness through online A/B tests. The experimental results reveal that \our{} achieves remarkable retrieval performance in widely used evaluation metrics. For instance, when implemented with the BGE-M3 model, \our{} elevates NDCG@10 from 73.96\% to 77.84\%, outperforming existing methods. A series of ablation studies coupled with pattern visualization further confirms that \our{} can effectively capture and utilize document structural semantics.

Our contributions can be summarized as follows. 
\begin{itemize}
    \item We propose \our, a novel contrastive learning framework explicitly incorporating document structural semantics through structure-aware learning and fine-grained element-level alignment, thereby enhancing structured data representations in a unified embedding space.
    \item We release \dataset, a dataset specifically designed for long structured document retrieval, which has over 10,000 words in document length on average and contains explicit structural information. 
    \item Extensive experiments across multiple modern PLMs demonstrate \our's consistent superiority over the state-of-the-art methods on both \dataset{} and industrial datasets. Online A/B testing further validates the effectiveness of \our.
\end{itemize}

\section{Related Work}

In this section, we review related work, including Pre-trained Language Models (PLMs), long document retrieval methods, and related benchmarks.

\subsection{Pre-trained Language Models}

The field of document retrieval has undergone transformative advancements driven by PLMs, particularly through the enhanced capability of extended context windows to effectively encode long documents. Seminal work by Karpukhin et al. \shortcite{Retrieval_open_qa} introduced dense passage retrieval for open-domain question answering, demonstrating the superior capabilities of PLMs in retrieval tasks. Subsequent research validated the effectiveness of BERT-based architectures \cite{modernBERT}, particularly through PLM integration in multi-stage document ranking \cite{Condenser}. Further innovations, such as ColBERT \cite{ColBERT}, advanced the field through contextualized late interaction mechanisms over BERT, enabling efficient passage retrieval. More recently, M3-Embedding \cite{M3_Embedding} has emerged as a state-of-the-art approach, leveraging self-knowledge distillation to optimize embedding quality and establish itself as a foundational architecture in retrieval systems. 

In contrast to the aforementioned encoder-only embedding models, decoder-only embedding models based on large language models (LLMs), such as gte-Qwen2-Instruct \cite{gte-Qwen2-Instruct}, MiniCPM-Embedding \cite{MiniCPM}, and NV-Embed \cite{NV-Embed}, introduce significantly higher latency (several times) during vector representation generation, substantially prolonging retrieval time. However, the resulting performance improvement is not commensurate with this considerable overhead. Consequently, this work primarily employs encoder-only pre-trained models.

\begin{figure*}[t]
\centering
\includegraphics[width=0.99\linewidth]{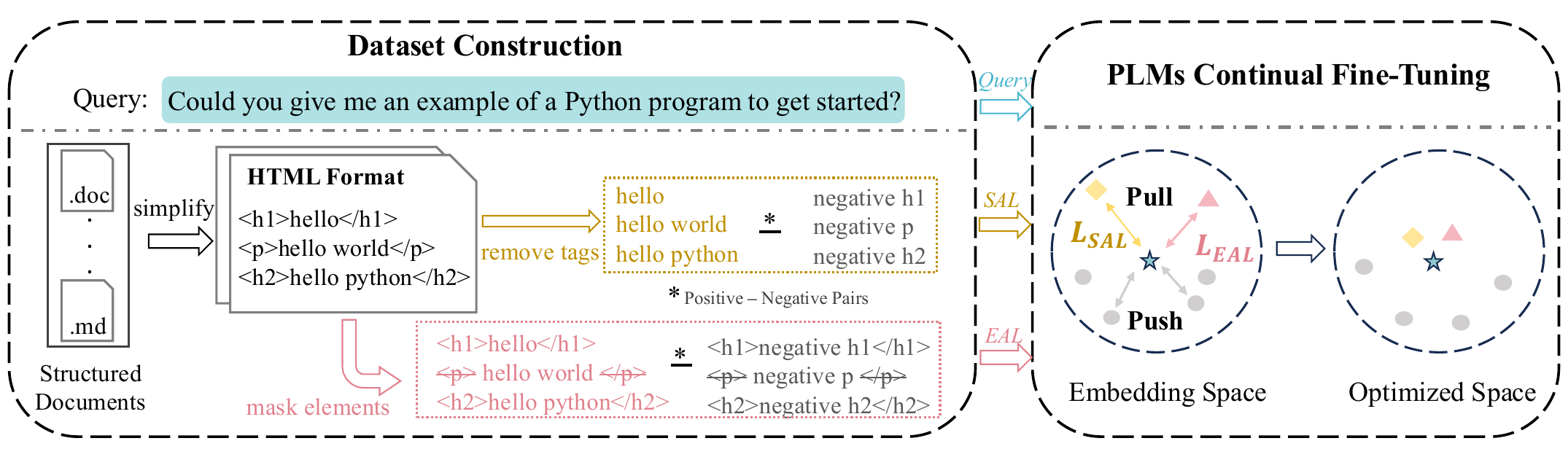} 
\caption{Framework of \our. We first construct the used dataset, including HTML transformation, tag processing, and element masking. To guide PLMs to map both queries and structured documents in a unified embedding space, we introduce Structure-Aware Learning (SAL) to incorporate document structural information and Element-Aware Alignment (EAL) to enhance semantic representation.}
\label{fig:overview}
\end{figure*} 

\subsection{Long Document Retrieval}

While contemporary document retrieval methods achieve remarkable performance in unstructured textual domains, their architectural limitations become apparent when handling structured data (e.g., technical specifications, legal instruments, and scholarly articles). Works in query optimization include ANCE \cite{ANCE} and DANCE \cite{DANCE}, which pioneer adaptive query expansion via contrastive dual learning, and Dai et al. \shortcite{EntailmentTuning}'s entailment tuning for dense passage retrieval such as open-domain question answering. 
Industrial-grade implementations like Facebook's EBR \cite{EBR} achieve scalability through hybrid embedding topologies, whereas MGDSPR \cite{MGDSPR} retrieves the most relevant products from a large corpus while retaining personalized user features in e-commerce retrieval. 
Longtriever \cite{Longtriever} divides long documents into short chunks and then models local semantics within the chunks and global context semantics between the chunks to improve retrieval.
Sun et al. \shortcite{sun2025zero} propose a hybrid retriever to obtain keyword and contextual information to further improve the quality of pseudo-documents.
SANTA \cite{SANTA} and CONAN \cite{CONAN} employ structure-aware pre-training protocols that combine structured data alignment and masked entity prediction for code retrieval and product retrieval.
However, these advances still exhibit a gap in addressing the retrieval requirements of long structured documents: insufficient capture of hierarchical structure and inability to perform fine-grained semantic alignment.

\subsection{Related Benchmarks}

Foundational benchmarks like MS MARCO \cite{MSMARCO}, TREC \cite{TREC}, and the multi-domain BEIR benchmark \cite{BEIR} have driven progress in supervised and zero-shot retrieval paradigms. Some datasets focus on specific tasks, such as product and code search \cite{ProductSearch, CodeSearch}. LongBench \cite{LongBench}, LongBench-V2 \cite{LongBenchV2}, and RULER \cite{RULER} are designed to assess LLM long-context reasoning, but are limited by the absence of document-grounded user queries and structural labels. Additionally, LongBench-V2 uses the choice format. Due to the above limitations, these works are insufficient for evaluating modern retrieval models in contemporary information retrieval systems.

\section{Methodology}

In this section, we first recall the preliminaries of long document retrieval. Subsequently, as shown in Figure \ref{fig:overview}, we introduce the framework of this work, including dataset construction, such as document pre-processing with tagging removal and element masking of structured documents, and continuous fine-tuning of PLMs with \our, which incorporates inherent structural features and fine-grained element alignment.

\paragraph{Preliminary of Document Retrieval}
Given a natural language query $q$ and a structured document corpus $D=\{d_i\}_{i=1}^n$, the retriever identifies the top-$k$ most relevant documents through a ranked list $\{d_1, d_2, ..., d_k\}$ of the $k$ most relevant documents, ranked by relevance scores.
Usually, we encode queries and structured documents with PLMs and map them into an embedding space for the calculation of the relevance score.

Let $\phi(\cdot) \in \mathbb{R}^l$ and $\varphi(\cdot) \in \mathbb{R}^l$ denote embedding functions that map queries and documents into an $l$-dimensional latent space, respectively. The relevance score of a query-document pair $(q,d)$ can be formally expressed as Equation~\ref{eq:sim_score} below.
\begin{equation}
    f(q, d) = sim (\phi(q), \varphi(d))
    \label{eq:sim_score}
\end{equation}
where $sim(\cdot)$ is a measurement function such as the inner product.

\subsection{Dataset Construction}

Given HTML's native hierarchical representation capabilities surpassing plain text in modeling retrieved knowledge \cite{html}, our method first converts structured documents into standardized HTML representations, where each constituent element preserves positional integrity via encapsulation within semantic markup tags. Subsequent pre-processing involves dual operations: (1) Tag Processing creates variants through retaining and removing tags, which facilitates the subsequent learning of basic structural features; (2) Element masking generates markup-depleted variants through stochastic tag elimination, enabling fine-grained alignment.

For industry data derived from real-world applications, we collect user-submitted queries through production system logs and acquire corresponding retrieved document lists via instrumentation in the engineering pipeline, with all documents stored in HTML format. 
For web-crawled documents, our pipeline initiates with targeted article acquisition, harvesting linked documents through breadth-first search crawling. Following data collection, we implement a preprocessing phase comprising removal of non-structural tags (e.g., line breaks and irrelevant markup tags) through regular expression pattern matching. The sanitized outputs subsequently undergo LLM-powered query synthesis to generate corresponding user queries. We designate this resource as StructDocRetrieval (MIT License). The details of the data example are shown in Table \ref{tab:data_example}.

\begin{table}[t]
\centering
\small
\begin{tabular}{p{0.18\columnwidth}|p{0.70\columnwidth}}
    \toprule
    \multicolumn{2}{l}{
    Query: How to use Python in VS Code?} \\ \midrule
    Relevant ~ Document & <title> [Nanny-level tutorial] VS Code installation and configuration of Python </title> <h1> Configure Jupyter in VS Code </h1> <h2> Install Jupyter extension </h2> <p> Choose the version that suits your computer and start downloading. </p> $\cdots$ \\ \midrule
    Irrelevant  ~ Document & <title> Data, algorithms, computing power, and blockchain + AI </title> <h1> Blockchain technology lays the foundation for a decentralized Internet. </h1> <p> Many investment firms have turned their attention to the emerging field of machine learning and artificial intelligence. </p> $\cdots$\\ \bottomrule
\end{tabular}
\caption{A data example of \dataset.}
\label{tab:data_example}
\end{table}

Table \ref{tab:data_statistic} presents statistics of the industrial dataset and StructDocRetrieval. Unlike typical short datasets, documents in our datasets are significantly longer, with an average of more than 7,000 words. In contrast, datasets like MS MARCO \cite{MSMARCO} have a maximum of 1,670 words, with most documents under 700 words. Furthermore, documents in MS MARCO and other variants are typically plain text, whereas StructDocRetrieval utilizes HTML format.

\begin{table}[t]
\centering
\small
\begin{tabular}{ccccc}
    \toprule 
     \multirow{2}{*}{Split} & \multirow{2}{*}{Query} & \multirow{2}{*}{Doc.} & \multicolumn{2}{c}{Avg. Words} \\
     & & & Query & Doc. \\
    \midrule
    \multicolumn{5}{c}{Industrial Dataset} \\
    Train & 12,047 & 8,580 & 10.07 & 7,310 \\
    Evaluation & 1,396 & 1,286 & 10.03 & 6,878 \\
    \midrule
    \multicolumn{5}{c}{\dataset} \\
    Train & 23816 & 23816 & 12.82 & 10,849 \\
    Test & 3404 & 3404 & 13.04 & 10,535 \\
    Evaluation & 6804 & 6804 & 12.74 & 11,047 \\
    \bottomrule
\end{tabular}
\caption{Data statistics of experiment datasets. The ``Avg. Words'' means the average number of words in queries and documents. 
}
\label{tab:data_statistic}
\end{table}

\subsection{Structure-Aware Learning}

PLMs have demonstrated remarkable capabilities in text representation learning via objectives like masked language modeling on large text corpora. However, their inherent lack of mechanisms to capture structural information hinders their ability to effectively comprehend and represent structured documents. This limitation consequently impacts the efficacy of structured document retrieval. To address this, we propose Structure-Aware Learning (SAL) to enhance PLMs with the capacity to encode structural information.

SAL aims to enable the model with structural awareness through a contrastive learning objective that leverages structural variants of relevant documents. We utilize preprocessed HTML-structured relevant documents as positive instances and irrelevant documents as negatives. To guide the model in recognizing the underlying structure, even without explicit tags, we derive plain text versions of these documents by removing all structural tags. The core idea is to train the model to distinguish query-aligned text originating from structured relevant documents from text originating from irrelevant documents.

As defined in Equation \ref{eq:sal}, the contrastive loss $\mathcal{L}_{SAL}$ maximizes the similarity between the query embedding $q$ and the embedding of the relevant document $d^+$, while minimizing the similarity with irrelevant documents $d^-$.
\begin{equation}
    \mathcal{L}_{SAL} = - log \frac{e^{f(q, d^+)}}{e^{f(q, d^+)} + \sum_{d^- \in D^-} e^{f(q, d^-)}}
    \label{eq:sal}
\end{equation}
The contrastive formulation incorporates dual variants: intact documents preserving markup semantics $d_{tag}^+$ and its destructured counterpart $d_{untag}^+$ with removed tags. The negative samples $D^- = \{d^-\}$ adopt the same process. 
Both tagged and untagged versions of $d^+$ and $d^-$ are used in the same loss computation. This design forces the model to recognize that structural semantics and the textual content of relevant documents should align with the query in the same embedding space.

\subsection{Element-Aware Alignment}

Masking strategies, such as masked language modeling \cite{MLM} and masked entity prediction \cite{mask_entiy, SANTA}, have proven effective in learning robust text representations. Unlike these approaches focusing on token or entity recovery, we introduce Element-Aware Alignment (EAL) based on masking structural elements within documents to foster fine-grained structural understanding.

For structured documents, we define elements as text spans annotated with tags (e.g., headings, list items, paragraphs). To encourage the model to learn fine-grained representations of these elements and their contextual roles, we randomly mask the structural tags of a proportion (e.g., 10\%) of elements in a document. Formally, let a structured document be represented as a sequence of elements $d = \{ (t_1, tag_1), (t_2, tag_2), \cdots, (t_n, tag_n) \}$, where $t_i$ is the text content and $tag_i$ is the structural tag.

A masked document $d_{mask}$ is constructed by removing $tag_i$ for a random subset of indices, while keeping other elements intact:
\begin{equation}
    d_{mask} = \{ \epsilon_1^{mask}, \epsilon_2, \epsilon_3^{mask}, \cdots, \epsilon_n \}
    \label{eq:masked_input}
\end{equation}
where $\epsilon_i^{mask}$ denotes the $i$-th element with its structural tag removed, while $\epsilon_i = (t_i, tag_i)$ preserves both the original text and its tag.

We form positive pairs using a query $q$ and its corresponding relevant document subjected to element masking $d_{mask}^+$. Negative pairs consist of $q$ and masked irrelevant documents $d_{mask}^-$. The training objective is based on a contrastive loss defined as follows:
\begin{equation}
    \mathcal{L}_{EAL} = - log \frac{e^{f(q, d_{mask}^+)}}{e^{f(q, d_{mask}^+)} + \sum_{d^- \in D^-} e^{f(q, d_{mask}^-)}}
    \label{eq:eal}
\end{equation}
Minimizing $\mathcal{L}_{EAL}$ trains the model to maintain high similarity between the query and the representation of the masked relevant document, while pushing away masked irrelevant documents. This objective forces the model to leverage the unmasked elements and the textual content within masked elements to infer the document’s relevance, thereby enhancing its ability to utilize fine-grained element-level information.

\begin{table*}[t]
\centering
\small
\begin{tabular}{c|ccccccc}
    \toprule 
    
    Method & 
    HitRate@1 &
    HitRate@3 &
    HitRate@5 &
    MRR@5 & 
    MRR@10 & 
    NDCG@5 & 
    NDCG@10 \\
    \midrule
    
    mE5-large & 54.11 & 79.62 & 85.86 & 67.39 & 68.06 & 72.18 & 74.11 \\
    \midrule
    + \textit{Chunk} & 56.85 & 82.94 & 88.79 & 70.12 & 71.45 & 74.78 & 77.42 \\
    + \textit{MCLS} & 57.74 & 84.12 & 89.56 & 71.08 & 72.41 & 75.76 & 78.44 \\
    + \textit{SANTA} & 55.79 & 81.76 & 88.02 & 69.01 & 70.49 & 73.79 & 76.50 \\
    + \textit{\our} & \textbf{58.63} & \textbf{85.29} & \textbf{90.34} & \textbf{72.02} & \textbf{73.37} & \textbf{76.74} & \textbf{79.35} \\
    
    \midrule
    
    bge-large-zh & 59.08 & 83.48 & 89.47 & 71.34 & 72.21 & 75.84 & 76.84 \\
    \midrule
    + \textit{Chunk} & 61.97 & 86.11 & 91.67 & 74.12 & 75.25 & 78.44 & 78.93 \\
    + \textit{MCLS} & 63.15 & 86.64 & 92.28 & 74.83 & 75.91 & 78.82 & 79.38 \\
    + \textit{SANTA} & 60.80 & 85.59 & 91.07 & 73.41 & 74.59 & 78.08 & 78.48 \\
    + \textit{\our} & \textbf{64.30} & \textbf{87.15} & \textbf{92.88} & \textbf{75.54} & \textbf{76.57} & \textbf{79.17} & \textbf{79.83} \\
    
    \midrule
    
    BGE-M3 & 61.03 & 85.24 & 91.69 & 73.35 & 73.96 & 77.97 & 79.41 \\
    \midrule
    + \textit{Chunk} & 64.28 & 87.69 & 93.19 & 76.11 & 76.91 & 80.76 & 81.52 \\
    + \textit{MCLS} & 65.27 & 87.98 & 93.48 & 76.74 & 77.37 & 81.14 & 82.05 \\
    + \textit{SANTA} & 63.27 & 87.42 & 92.91 & 75.48 & 76.44 & 80.38 & 81.00 \\
    + \textit{\our} & \textbf{66.26} & \textbf{88.25} & \textbf{93.77} & \textbf{77.38} & \textbf{77.84} & \textbf{81.52} & \textbf{82.59} \\

    \bottomrule
\end{tabular}
\caption{Retrieval effectiveness of different models on the industrial dataset.}
\label{tab:main_results}
\end{table*}

\section{Experiments}

In this section, we evaluate \our{} across various datasets using different PLMs. We further present in-depth studies of \our, including ablation studies and embedding distributions visualization. Additionally, we show the practical improvement of \our{} in the industrial environment.

\subsection{Setup}
We describe the basic experiment setup used in our work in this section.

\paragraph{Datasets and Models}
This study utilizes industry data from real-world applications.
User-clicked documents serve as positive examples (target documents), while non-clicked documents are treated as negative examples. All documents are stored in HTML format. 
We select a set of modern embedding models as baselines, including Multilingual-E5-large \cite{wang2024multilingual}, bge-large-zh \cite{bge_embedding}, BGE-M3 \cite{M3_Embedding}, and GTE-Qwen2-1.5B \cite{gte-Qwen2-Instruct}.

\paragraph{Evaluation Metrics}
We adopt three widely-used metrics: HitRate, MRR, and NDCG. 
HitRate reflects immediate retrieval accuracy, MRR emphasizes the ability of the model to prioritize critical items, and NDCG considers graded relevance and positional sensitivity.

\paragraph{Baselines}

We compare \our{} with state-of-the-art document retrieval methods: Chunk, MCLS \cite{M3_Embedding}, and SANTA\cite{SANTA}. 

Chunk-based processing is a conventional solution in document retrieval. Long documents are segmented into fixed-length chunks of 512 tokens, each independently encoded via PLMs. Query-document relevance is determined by computing dot product similarities between the query embedding and each chunk's embedding, with the maximum value retained as the final score.

For MCLS, we insert a ``[CLS]'' token for every fixed number of tokens (inserting a [CLS] token for each 256 tokens in our experiments). The final document embedding is computed by averaging the last hidden states of all [CLS] tokens.

To adapt Masked Entity Prediction of SANTA for the latest encoder-only models like BGE-M3, we use the same tool to identify co-occurring terms in the Query, Title, and Body as entities, apply random masking, and utilize the model to predict the masked entity tokens.

\paragraph{Implementations}

We begin with the fine-tuning of contrastive learning of the base PLMs. All the methods demonstrate performance improvements over the fine-tuned PLMs. We use the Adam optimizer with a learning rate of 1e-5 and 2 training epochs. The maximum query length is 32, and the maximum sequence length is 4096. We sample 8 negative samples for each query and use cross device negatives, the total batch size is 8. All the implementations utilize PyTorch and FlagEmbedding\footnote{https://github.com/FlagOpen/FlagEmbedding}. The experiments are performed on 4 NVIDIA A800 GPUs.

\begin{table}[t]
\centering
\small
\begin{tabular}{c|ccc}
    \toprule 
    
    Method & 
    HitRate@5 &
    MRR@10 & 
    NDCG@10 \\
    \midrule
    mE5-large & 92.89 & 83.02 & 86.24 \\
    \midrule
    + \textit{Chunk} & 93.95 & 84.90 & 87.67 \\
    + \textit{MCLS} & 94.16 & 85.39 & 88.38 \\
    + \textit{SANTA} & 93.58 & 84.31 & 87.31 \\
    + \textit{\our} & \textbf{94.72} & \textbf{86.53} & \textbf{89.31} \\ 
    \midrule
    bge-large-zh & 94.59 & 85.95 & 88.68 \\
    \midrule
    + \textit{Chunk} & 95.65 & 87.83 & 90.11 \\
    + \textit{MCLS} & 95.86 & 88.32 & 90.82 \\
    + \textit{SANTA} & 95.28 & 87.24 & 89.75 \\
    + \textit{\our} & \textbf{96.42} & \textbf{89.46} & \textbf{91.75} \\
    \midrule
    BGE-M3 & 95.39 & 87.36 & 89.92 \\
    \midrule
    + \textit{Chunk} & 96.45 & 89.24 & 91.35 \\
    + \textit{MCLS} & 96.82 & 89.78 & 91.95 \\
    + \textit{SANTA} & 95.98 & 88.45 & 90.83 \\
    + \textit{\our} & \textbf{97.09} & \textbf{90.10} & \textbf{92.25} \\
    \bottomrule
\end{tabular}
\caption{The retrieval performance on \dataset.} 
\label{tab:web}
\end{table}

\subsection{Overall Performance}
Retrieval effectiveness is evaluated on two distinct datasets: a real industrial dataset and the \dataset{} web dataset, utilizing three different base embedding models (mE5-large, bge-large-zh, and BGE-M3). We summarize the performance results across various standard metrics, including Hitrate@k, MRR@k, and NDCG@k.

Table \ref{tab:main_results} presents comparative retrieval effectiveness across real industrial structured documents, where \our{} achieves the state-of-the-art performance with HitRate@3 absolute gains of 5.67\% over fine-tuned baselines and 3.53\% over existing structural-aware methods. This performance advantage persists in web-crawled retrieval dataset \dataset{} (cf. Table \ref{tab:web}), particularly in high-recall metrics (i.e., HitRate@5: +1.70\% avg.) and precision-sensitive measures (i.e., NDCG@10: +2.33\% avg.). The consistent improvements across both controlled industrial and diverse web environments validate that \our{} enables the advantages of PLMs in representing long structured documents, making PLMs sensitive to document structures and better at representing structured data.

\subsection{In-depth Analysis}

In this subsection, we present an in-depth analysis, including ablation studies to investigate the contributions of two fundamental components of our method, an investigation of mask ratios in element-aware alignment, a comparison of training strategies, visualizations of the learned embedding distributions, and validation with an extended-context model.

\paragraph{Ablation Study}

In this work, we employ structure-aware learning (SAL) and element-aware alignment (EAL) to conduct continuous training on the BGE-M3 model, demonstrating their effectiveness in guiding the model to better learn semantic features from structured documents. Table \ref{tab:ablation} presents the retrieval performance of these two fundamental components.

Compared to the baseline model, SAL and EAL exhibit divergent performance in structured data retrieval tasks. SAL shows no significant improvement over the baseline, restricted by the dependence on tag awareness alone to distinguish structured documents. In contrast, EAL achieves substantial enhancements in all the evaluation metrics.
The superiority of EAL is due to its contrastive training paradigm between structural elements (e.g., HTML tags) and unstructured text components. This methodology effectively bridges the modality gap between heterogeneous data types through joint embedding space projection, thereby facilitating cross-modal representation learning with enhanced retrieval efficacy.

Overall, integrating augmented tasks through SAL and EAL yields progressive performance gains. These empirical results confirm that explicit structural awareness enables models to better encode semantic hierarchies while generating optimized textual representations tailored for complex structured data environments.

\begin{table}[t]
\centering
\small
\begin{tabular}{c|ccc}
    \toprule 
    
    Method & 
    HitRate@5 &
    MRR@10 & 
    NDCG@10 \\
    \midrule
    BGE-M3 & 91.69 & 73.96 & 79.41 \\
    w/ $SAL$ & 91.98 & 74.69  & 80.08 \\
    w/ $EAL$ & 92.12 & 75.83 & 80.85 \\
    w/ $\our$ & \textbf{93.77} & \textbf{77.84} & \textbf{82.59} \\
    \bottomrule
\end{tabular}
\caption{The performance of ablation models on industrial structured document retrieval.} 
\label{tab:ablation}
\end{table}

\begin{table}[t]
\centering
\small
\begin{tabular}{c|ccc}
    \toprule 
    
    ratios (\%) & 
    HitRate@5 &
    MRR@10 & 
    NDCG@10 \\

    \midrule
    1 & 86.71 & 75.13 & 80.67 \\
    5 & 93.24 & 77.11 & 81.98 \\
    10 & \textbf{93.77} & \textbf{77.84} & \textbf{82.59} \\
    30 & 93.17 & 76.95 & 82.31 \\
    50 & 92.99 & 76.66 & 81.67 \\
    
    \bottomrule
\end{tabular}
\caption{The impact of element-aware alignment mask ratios on BGE-M3.}
\label{tab:mask_ratios}
\end{table}

\paragraph{Impact of Mask Ratios}

We further investigate the impact of different element mask ratios in the element-aware alignment on retrieval effectiveness to determine the optimal configuration.
Five distinct mask ratios (1\%, 5\%, 10\%, 30\%, and 50\%) are evaluated using the BGE-M3 model with performance metrics including HitRate@5, NDCG@10, and MRR@10. The Experimental results are shown in Table~\ref{tab:mask_ratios}. 
It can be observed that (1) the 10\% mask ratio achieves optimal performance across all the metrics, and (2) performance variations remain marginal across different ratios, such as NDCG@10 differences constrained within a 2\% range, demonstrating the robustness of our method to mask ratio selection.
Based on these findings, we adopt the 10\% masking ratio as the default configuration in the experimental section.

\paragraph{Impact of Training Strategy}

During continual fine-tuning, our empirical analysis reveals that simply combining $L_{EAL}$ and $L_{SAL}$ as $\mathcal{L} = (\mathcal{L}_{SAL} + \mathcal{L}_{EAL})$ does not achieve the best performance. Therefore, we make explorations for the impacts of different optimization strategies. 
As shown in Table \ref{tab:training}, $EAL^1-SAL^2$ achieves superior performance. The superiority stems from: (1) Local Semantics Foundation: $EAL$ learns local semantic sensitivity by randomly masking element tags (e.g., <h1>), providing a high-quality textual representation for subsequent structural understanding. (2) Progressive Difficulty: $EAL$ aligns local elements with queries (simpler task), while SAL integrates global structure with query intent (complex task).  The $EAL \rightarrow SAL$ sequence follows an easy-to-hard trajectory, thereby preventing premature overfitting to structural noise.

\begin{table}[t]
\centering
\small
\begin{tabular}{c|ccc}
    \toprule 
    
    Method & 
    HitRate@5 &
    MRR@10 & 
    NDCG@10 \\
    
    \midrule
    
    BGE-M3 & 91.69 & 73.96 & 79.41 \\
    $SAL-EAL$ & 91.76 & 74.73 & 80.06 \\
    $SAL^1-EAL^2$ & 93.27 & 76.98 & 81.96 \\
    $EAL^1-SAL^2$ & \textbf{93.77} & \textbf{77.84} & \textbf{82.59} \\
    
    \bottomrule
\end{tabular}
\caption{The retrieval performance of different training strategies.
The ``$SAL-EAL$'' means performing structure-aware and element-aware learning simultaneously. 
``$SAL^1-EAL^2$'' indicates that structure-aware learning is performed first, followed by element-aware alignment, while ``$EAL^1-SAL^2$'' is the opposite.
}
\label{tab:training}
\end{table}

\paragraph{Visualization}

\begin{figure}[t]
\centering 
    \subfigure[BGE-M3]{
        \includegraphics[width=0.225\textwidth]{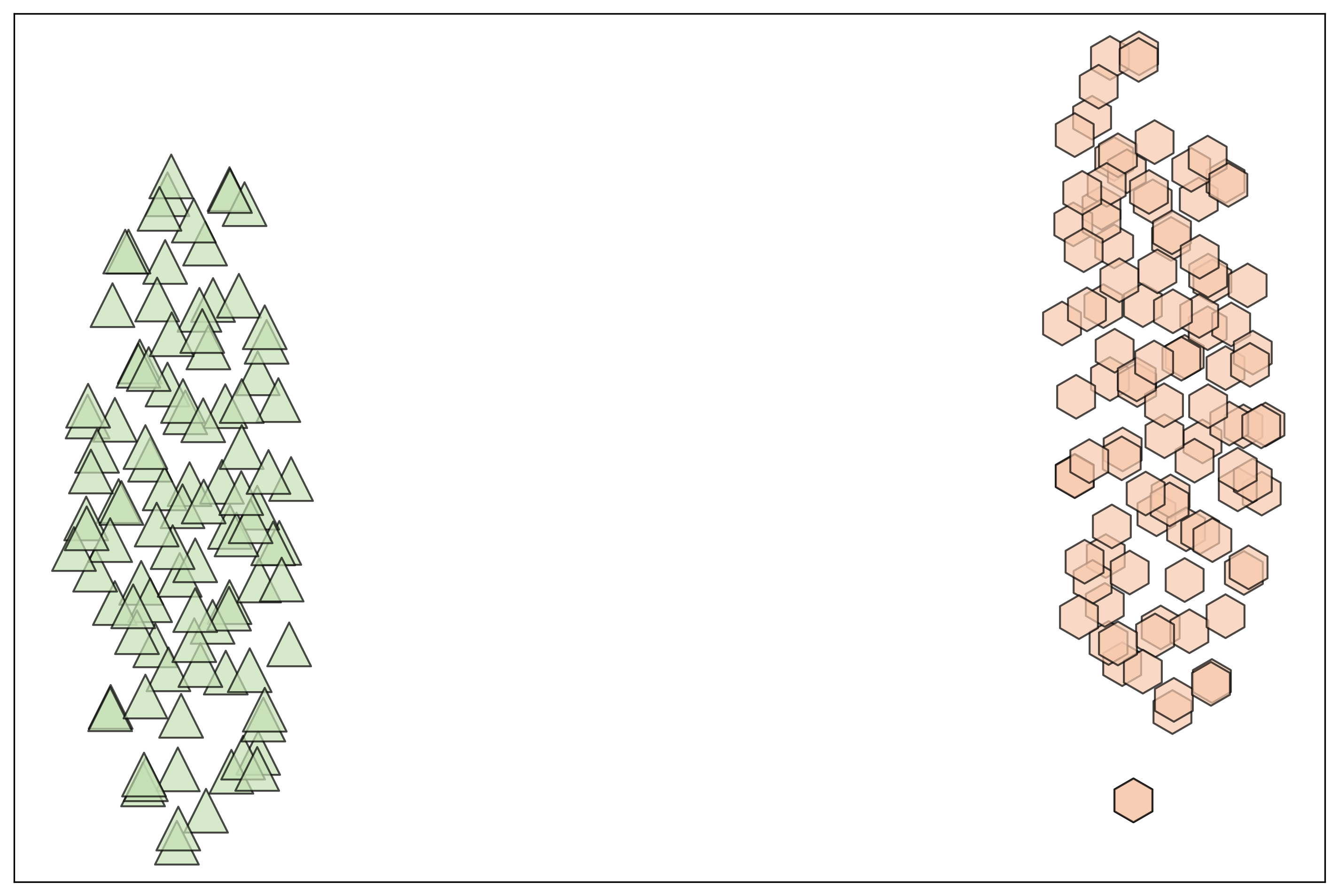}
    }
    \subfigure[BGE-M3 w/ SAL]{
        \includegraphics[width=0.225\textwidth]{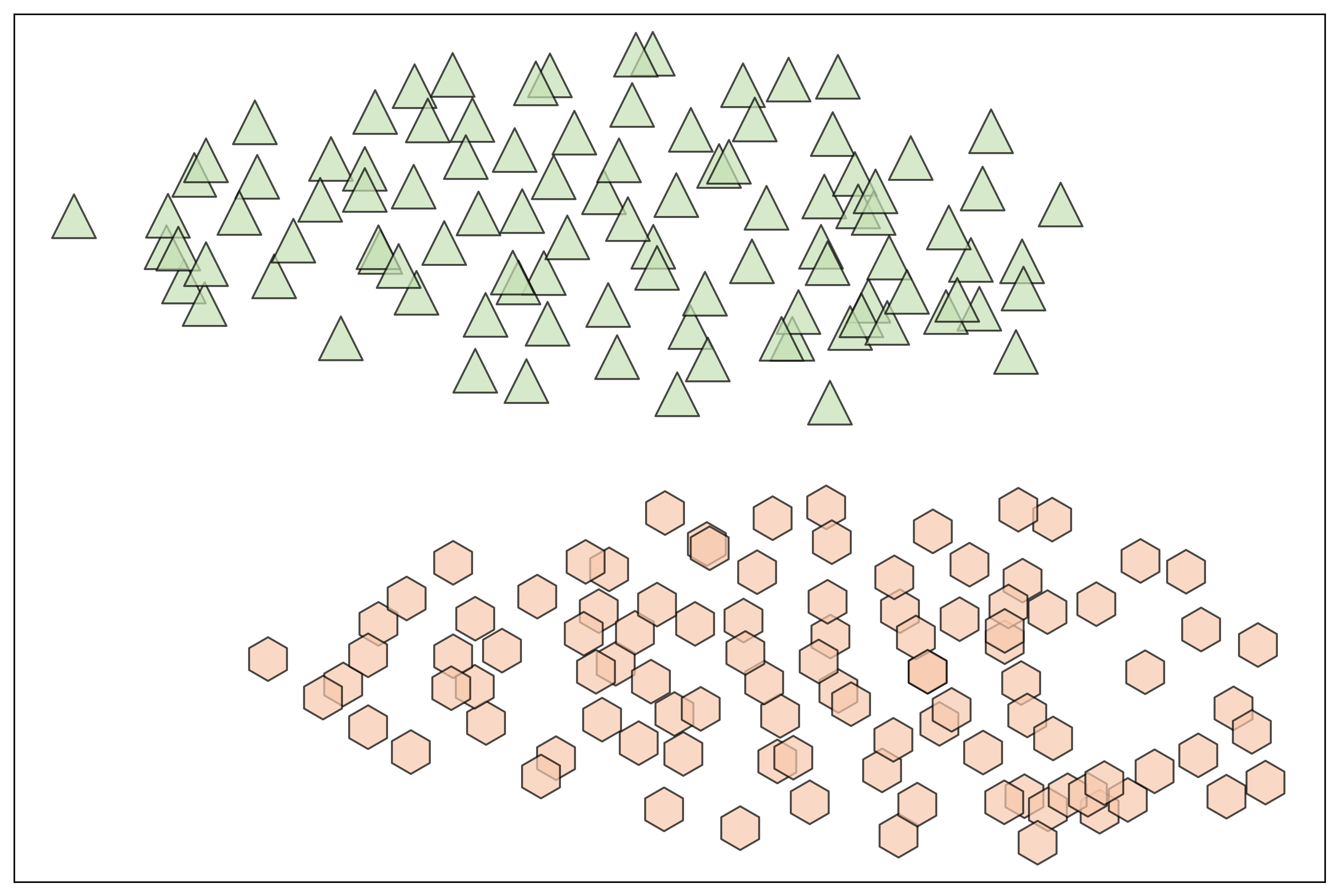}
    }
    \subfigure[BGE-M3 w/ EAL]{
        \includegraphics[width=0.225\textwidth]{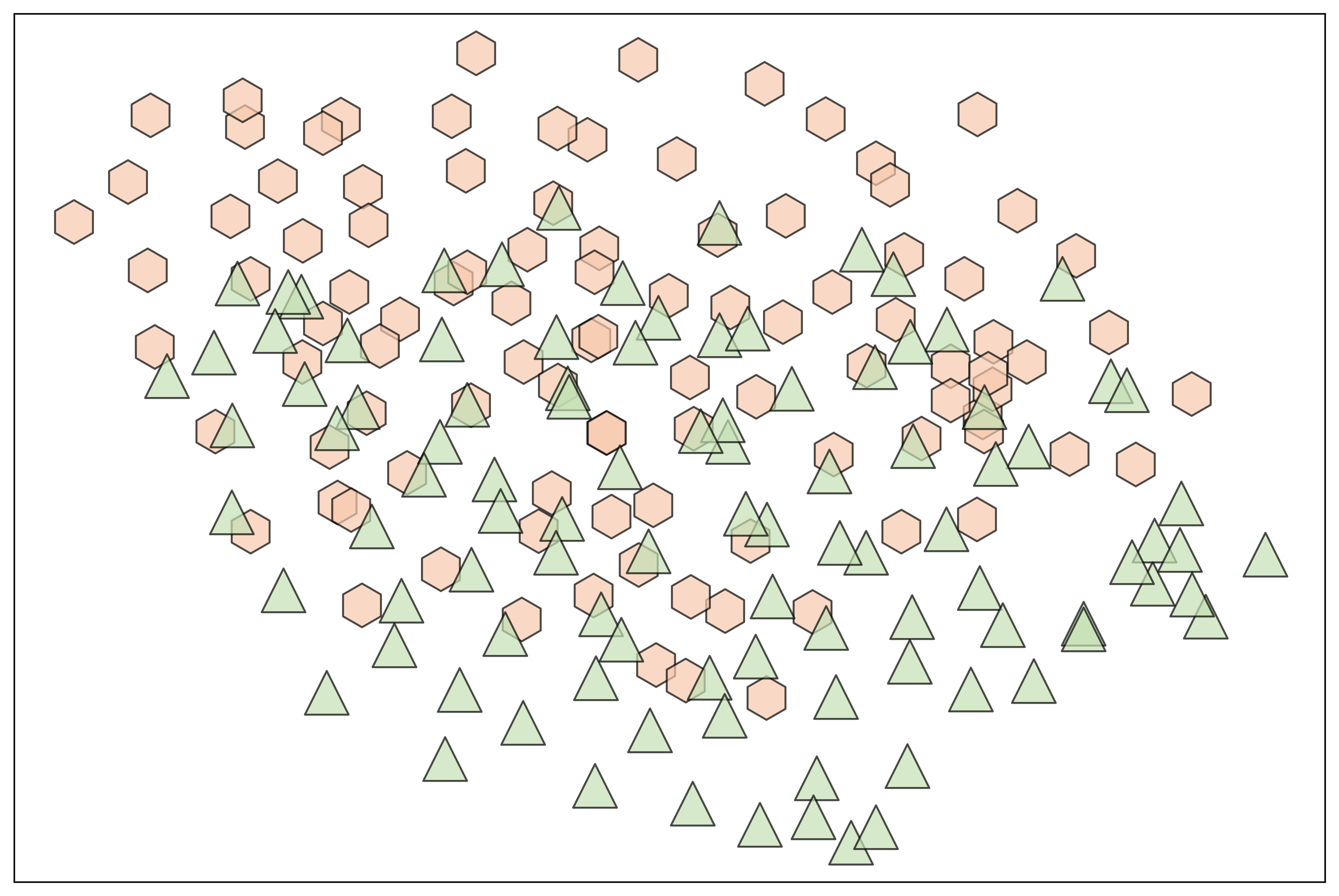}
    }
    \subfigure[BGE-M3 w/ SEAL]{
        \includegraphics[width=0.225\textwidth]{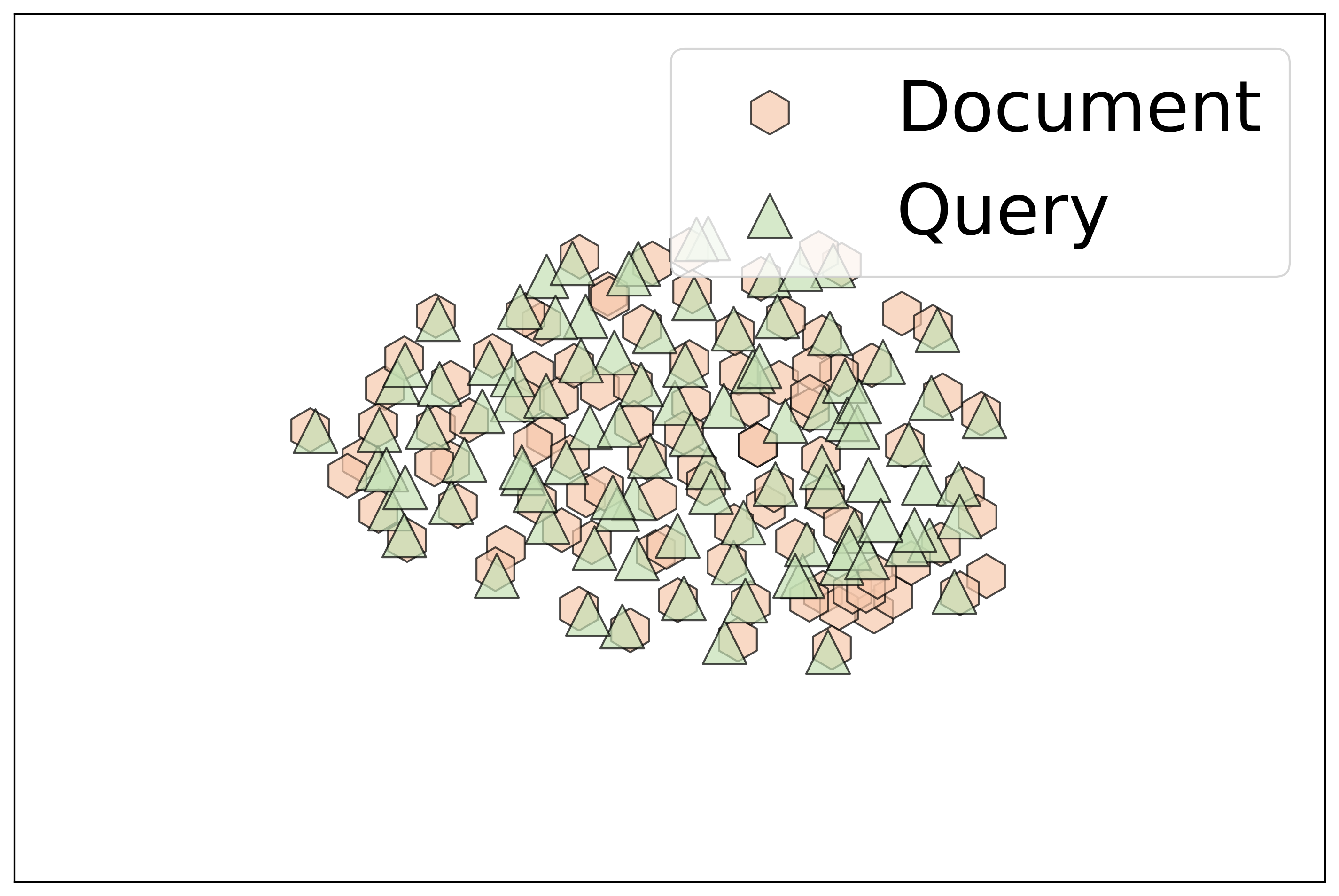}
    }
\caption{Embedding visualization of original model and \our{} using T-SNE.}
\label{fig:visual}
\end{figure}

To evaluate the quality of the learned representations and validate the establishment of a unified embedding space for queries and structured documents, we visualize the latent representations of queries and their corresponding documentation texts in Figure \ref{fig:visual}. An ablation study is also presented to isolate the contributions of Structure-Aware Learning and Element-Aware Alignment.

Comparative analysis between Figure \ref{fig:visual} (a) and \ref{fig:visual} (b) reveals that the effect of incorporating SAL. This integration reduces the divergence between query and documentation embeddings, demonstrating enhanced capture of structural context. Figure \ref{fig:visual} (c) illustrates that EAL significantly improves query-document alignment. This improvement is evidenced by increased semantic proximity and tighter cluster fusion between queries and documents in the latent space. This suggests that our fine-grained masked element alignment mechanism enhances the model's capacity to capture more specific semantic distinctions for the refined embedding space.

Finally, comparison of Figures \ref{fig:visual} (a) and \ref{fig:visual} (d) indicates that \our{} exhibits superior embedding homogeneity compared to the base model through its dual mechanisms of structure-aware learning and fine-grained element alignment. These visualizations demonstrate that \our{} effectively learns a more unified and well-structured representation learning for structured document retrieval.

\paragraph{Extended-context model}

The experimental validity is strengthened through robustness testing with extended-context models. We introduce GTE-Qwen2-1.5B (32k-token context window) as an extended baseline, demonstrating that: (1) observed performance improvements are consistent with our previous results; (2) model efficacy exhibits progressive scalability across varying context window specifications.

\begin{figure}[t]
\centering
\includegraphics[width=0.99\columnwidth]{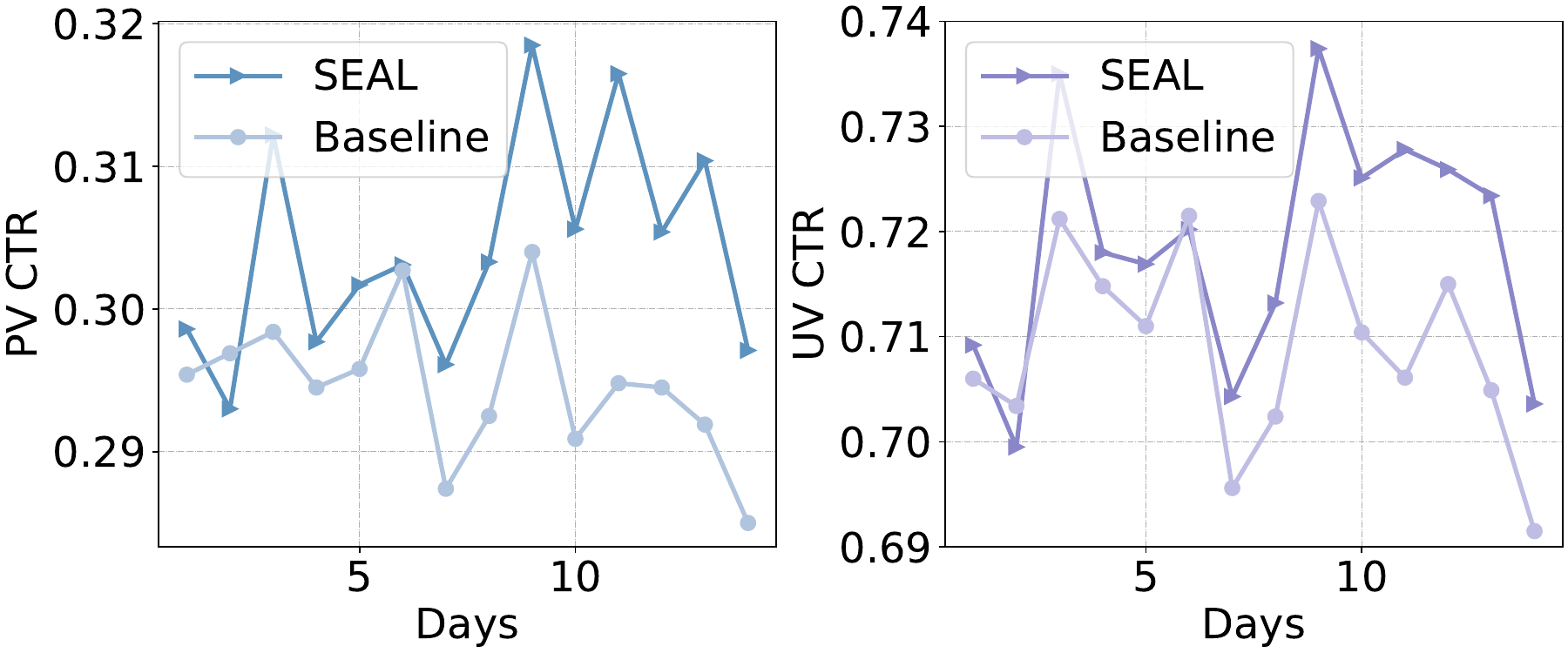} 
\caption{Online PV CTR and UV CTR over a two-Week period.}
\label{fig:ABTest}
\end{figure}

\begin{table}[t]
\centering
\small
\begin{tabular}{c|ccc}
    \toprule 
    
    Method & 
    HitRate@5 &
    MRR@10 & 
    NDCG@10 \\
    \midrule
    mE5-large & 92.17 & 74.32 & 79.86 \\
    \midrule
    + \textit{Chunk} & 93.25 & 77.18 & 81.68 \\
    + \textit{MCLS} & 94.00 & 77.51 & 82.18 \\
    + \textit{\our} & \textbf{94.72} & \textbf{78.17} & \textbf{83.56} \\ 
    \midrule
    
\end{tabular}
\caption{The retrieval performance of GTE-Qwen2-1.5B on \dataset.}
\label{tab:qwen}
\end{table}

\subsection{Online A/B Testing}

\our{} has been deployed in the practical platform for long-document retrieval services with hundreds of thousands of daily active users. We compare the performance of \our{} and a baseline method that performs raw-text contrastive learning over a 14-day period, evaluated using both PV CTR (Page View Click-Through Rate, clicks of ads divided by impressions of ads views) and UV CTR (Unique Visitor Click-Through Rate, clicks of unique visitors divided by impressions of unique visitors). The former is used to evaluate how often a page is clicked when it is browsed, and the latter reflects the attractiveness of the page to users.

As shown in Figure \ref{fig:ABTest}, compared to the previously deployed model, our approach demonstrates performance improvements in online A/B testing. The experiment, which accounted for approximately 30\% of search traffic, was conducted over a two-week period. \our{} achieves an average improvement of 1.6\% and 1.2\% in PV CTR and UV CTR without introducing additional overhead, and higher PV and UV CTR on 12 out of 14 days. The superior performance of \our{} in both metrics demonstrates its effective optimization of both page-level attractiveness and user-level engagement, ultimately enhancing the user experience in practical applications.

\section{Conclusion}

In this work, we propose \our, a novel contrastive framework that integrates structural awareness and fine-grained semantic alignment to enhance PLMs for long structured document retrieval.
To foster research in this field, we release StructDocRetrieval, a dataset of long documents enriched with structural information, establishing an evaluation scenario close to real-world applications.
Extensive experiments demonstrate \our{} achieves state-of-the-art performance in long structured document retrieval across various PLMs and datasets. Our in-depth analysis further reveals that \our{} induces a unified embedding space that effectively aligns queries and relevant documents.

\section*{Limitations}

Although \our{} exhibits strong performance in long structured document retrieval, its reliance on alignment signals between structured and unstructured data raises open questions about its general superiority over baseline models in all downstream tasks, such as code retrieval. Additionally, our current empirical validation focuses primarily on the Chinese-language community, though we are actively constructing an English-language corpus to assess generalization capabilities. Finally, our approach preserves the potential of exploiting the document structure during pre-training.

\section*{Acknowledgments}
This work is supported by the following funding sources: the Guangzhou Industrial Information and Intelligent Key Laboratory Project (No.2024A03J0628); the National Natural Science Foundation of China (NSFC) (No.62306256); the Natural Science Foundation of Guangdong Province (No.2025A1515010261); and the Key R\&D Program of Zhejiang Province (No. 2024C01036).

\bibliography{SEAL}

\begin{thebibliography}{40}
\providecommand{\natexlab}[1]{#1}

\bibitem[{Bai et~al.(2024{\natexlab{a}})Bai, Lv, Zhang, Lyu, Tang, Huang, Du,
  Liu, Zeng, Hou, Dong, Tang, and Li}]{LongBench}
Yushi Bai, Xin Lv, Jiajie Zhang, Hongchang Lyu, Jiankai Tang, Zhidian Huang,
  Zhengxiao Du, Xiao Liu, Aohan Zeng, Lei Hou, Yuxiao Dong, Jie Tang, and
  Juanzi Li. 2024{\natexlab{a}}.
\newblock \href {https://doi.org/10.18653/v1/2024.acl-long.172} {Longbench: {A}
  bilingual, multitask benchmark for long context understanding}.
\newblock In \emph{{ACL} {(1)}}, pages 3119--3137. Association for
  Computational Linguistics.

\bibitem[{Bai et~al.(2024{\natexlab{b}})Bai, Tu, Zhang, Peng, Wang, Lv, Cao,
  Xu, Hou, Dong, Tang, and Li}]{LongBenchV2}
Yushi Bai, Shangqing Tu, Jiajie Zhang, Hao Peng, Xiaozhi Wang, Xin Lv, Shulin
  Cao, Jiazheng Xu, Lei Hou, Yuxiao Dong, Jie Tang, and Juanzi Li.
  2024{\natexlab{b}}.
\newblock \href {https://doi.org/10.48550/arXiv.2412.15204} {Longbench v2:
  Towards deeper understanding and reasoning on realistic long-context
  multitasks}.
\newblock \emph{CoRR}, abs/2412.15204.

\bibitem[{Chen et~al.(2024)Chen, Xiao, Zhang, Luo, Lian, and
  Liu}]{M3_Embedding}
Jianlyu Chen, Shitao Xiao, Peitian Zhang, Kun Luo, Defu Lian, and Zheng Liu.
  2024.
\newblock \href {https://doi.org/10.18653/v1/2024.findings-acl.137}
  {M3-embedding: Multi-linguality, multi-functionality, multi-granularity text
  embeddings through self-knowledge distillation}.
\newblock In \emph{{ACL} (Findings)}, pages 2318--2335. Association for
  Computational Linguistics.

\bibitem[{Craswell et~al.(2020)Craswell, Mitra, Yilmaz, Campos, and
  Voorhees}]{TREC}
Nick Craswell, Bhaskar Mitra, Emine Yilmaz, Daniel Campos, and Ellen~M.
  Voorhees. 2020.
\newblock Overview of the {TREC} 2019 deep learning track.
\newblock \emph{CoRR}, abs/2003.07820.

\bibitem[{Dai et~al.(2024)Dai, Liu, and Xiong}]{EntailmentTuning}
Lu~Dai, Hao Liu, and Hui Xiong. 2024.
\newblock \href {https://aclanthology.org/2024.emnlp-main.636} {Improve dense
  passage retrieval with entailment tuning}.
\newblock In \emph{{EMNLP} {(1)}}. Association for Computational Linguistics.

\bibitem[{Devlin et~al.(2019)Devlin, Chang, Lee, and Toutanova}]{BERT}
Jacob Devlin, Ming{-}Wei Chang, Kenton Lee, and Kristina Toutanova. 2019.
\newblock \href {https://doi.org/10.18653/v1/n19-1423} {{BERT:} pre-training of
  deep bidirectional transformers for language understanding}.
\newblock In \emph{{NAACL-HLT} {(1)}}, pages 4171--4186. Association for
  Computational Linguistics.

\bibitem[{Gao and Callan(2021)}]{Condenser}
Luyu Gao and Jamie Callan. 2021.
\newblock \href {https://doi.org/10.18653/v1/2020.emnlp-main.550} {Condenser: a
  pre-training architecture for dense retrieval}.
\newblock In \emph{{EMNLP} {(1)}}, pages 981--993. Association for
  Computational Linguistics.

\bibitem[{Gupta et~al.(2024)Gupta, Ranjan, and Singh}]{RAG_Survey}
Shailja Gupta, Rajesh Ranjan, and Surya~Narayan Singh. 2024.
\newblock \href {https://doi.org/10.48550/arXiv.2410.12837} {A comprehensive
  survey of retrieval-augmented generation {(RAG):} evolution, current
  landscape and future directions}.
\newblock \emph{CoRR}, abs/2410.12837.

\bibitem[{Hsieh et~al.(2024)Hsieh, Sun, Kriman, Acharya, Rekesh, Jia, Zhang,
  and Ginsburg}]{RULER}
Cheng{-}Ping Hsieh, Simeng Sun, Samuel Kriman, Shantanu Acharya, Dima Rekesh,
  Fei Jia, Yang Zhang, and Boris Ginsburg. 2024.
\newblock \href {https://doi.org/10.48550/arXiv.2404.06654} {{RULER:} what's
  the real context size of your long-context language models?}
\newblock \emph{CoRR}, abs/2404.06654.

\bibitem[{Hu et~al.(2024)Hu, Tu, Han, He, Cui, Long, Zheng, Fang, Huang, Zhao,
  Zhang, Thai, Zhang, Wang, Yao, Zhao, Zhou, Cai, Zhai, Ding, Jia, Zeng, Li,
  Liu, and Sun}]{MiniCPM}
Shengding Hu, Yuge Tu, Xu~Han, Chaoqun He, Ganqu Cui, Xiang Long, Zhi Zheng,
  Yewei Fang, Yuxiang Huang, Weilin Zhao, Xinrong Zhang, Zhen~Leng Thai, Kai
  Zhang, Chongyi Wang, Yuan Yao, Chenyang Zhao, Jie Zhou, Jie Cai, Zhongwu
  Zhai, Ning Ding, Chao Jia, Guoyang Zeng, Dahai Li, Zhiyuan Liu, and Maosong
  Sun. 2024.
\newblock \href {https://doi.org/10.48550/arXiv.2404.06395} {Minicpm: Unveiling
  the potential of small language models with scalable training strategies}.
\newblock \emph{CoRR}, abs/2404.06395.

\bibitem[{Huang et~al.(2020)Huang, Sharma, Sun, Xia, Zhang, Pronin,
  Padmanabhan, Ottaviano, and Yang}]{EBR}
Jui{-}Ting Huang, Ashish Sharma, Shuying Sun, Li~Xia, David Zhang, Philip
  Pronin, Janani Padmanabhan, Giuseppe Ottaviano, and Linjun Yang. 2020.
\newblock \href {https://doi.org/10.1145/3394486.3403305} {Embedding-based
  retrieval in facebook search}.
\newblock In \emph{{KDD}}, pages 2553--2561. {ACM}.

\bibitem[{Husain et~al.(2019)Husain, Wu, Gazit, Allamanis, and
  Brockschmidt}]{CodeSearch}
Hamel Husain, Ho{-}Hsiang Wu, Tiferet Gazit, Miltiadis Allamanis, and Marc
  Brockschmidt. 2019.
\newblock \href {http://arxiv.org/abs/1909.09436} {Codesearchnet challenge:
  Evaluating the state of semantic code search}.
\newblock \emph{CoRR}, abs/1909.09436.

\bibitem[{Karpukhin et~al.(2020{\natexlab{a}})Karpukhin, Oguz, Min, Lewis, Wu,
  Edunov, Chen, and Yih}]{PassageRetrieval}
Vladimir Karpukhin, Barlas Oguz, Sewon Min, Patrick Lewis, Ledell Wu, Sergey
  Edunov, Danqi Chen, and Wen{-}tau Yih. 2020{\natexlab{a}}.
\newblock \href {https://doi.org/10.18653/v1/2020.emnlp-main.550} {Dense
  passage retrieval for open-domain question answering}.
\newblock In \emph{{EMNLP} {(1)}}, pages 6769--6781. Association for
  Computational Linguistics.

\bibitem[{Karpukhin et~al.(2020{\natexlab{b}})Karpukhin, Oguz, Min, Lewis, Wu,
  Edunov, Chen, and Yih}]{Retrieval_open_qa}
Vladimir Karpukhin, Barlas Oguz, Sewon Min, Patrick S.~H. Lewis, Ledell Wu,
  Sergey Edunov, Danqi Chen, and Wen{-}tau Yih. 2020{\natexlab{b}}.
\newblock \href {https://doi.org/10.18653/v1/2020.emnlp-main.550} {Dense
  passage retrieval for open-domain question answering}.
\newblock In \emph{{EMNLP} {(1)}}, pages 6769--6781. Association for
  Computational Linguistics.

\bibitem[{Khattab and Zaharia(2020)}]{ColBERT}
Omar Khattab and Matei Zaharia. 2020.
\newblock \href {https://doi.org/10.1145/3397271.3401075} {Colbert: Efficient
  and effective passage search via contextualized late interaction over
  {BERT}}.
\newblock In \emph{{SIGIR}}, pages 39--48. {ACM}.

\bibitem[{Lee et~al.(2025)Lee, Roy, Xu, Raiman, Shoeybi, Catanzaro, and
  Ping}]{NV-Embed}
Chankyu Lee, Rajarshi Roy, Mengyao Xu, Jonathan Raiman, Mohammad Shoeybi, Bryan
  Catanzaro, and Wei Ping. 2025.
\newblock \href {https://openreview.net/forum?id=lgsyLSsDRe} {Nv-embed:
  Improved techniques for training llms as generalist embedding models}.
\newblock In \emph{{ICLR}}. OpenReview.net.

\bibitem[{Li et~al.(2020)Li, Zhou, He, Wang, Yang, and Li}]{MLM}
Bohan Li, Hao Zhou, Junxian He, Mingxuan Wang, Yiming Yang, and Lei Li. 2020.
\newblock \href {https://doi.org/10.18653/v1/2020.emnlp-main.733} {On the
  sentence embeddings from pre-trained language models}.
\newblock In \emph{{EMNLP} {(1)}}, pages 9119--9130. Association for
  Computational Linguistics.

\bibitem[{Li et~al.(2021{\natexlab{a}})Li, Lv, Jin, Lin, Yang, Zeng, Wu, and
  Ma}]{MGDSPR}
Sen Li, Fuyu Lv, Taiwei Jin, Guli Lin, Keping Yang, Xiaoyi Zeng, Xiao{-}Ming
  Wu, and Qianli Ma. 2021{\natexlab{a}}.
\newblock \href {https://doi.org/10.1145/3447548.3467101} {Embedding-based
  product retrieval in taobao search}.
\newblock In \emph{{KDD}}, pages 3181--3189. {ACM}.

\bibitem[{Li et~al.(2022)Li, Gong, Shen, Qiu, Zhang, Yao, Qi, Jiang, Chen, and
  Duan}]{CodeRetriever}
Xiaonan Li, Yeyun Gong, Yelong Shen, Xipeng Qiu, Hang Zhang, Bolun Yao, Weizhen
  Qi, Daxin Jiang, Weizhu Chen, and Nan Duan. 2022.
\newblock \href {https://doi.org/10.18653/v1/2022.emnlp-main.187}
  {{C}ode{R}etriever: A large scale contrastive pre-training method for code
  search}.
\newblock In \emph{{ACL}}, pages 2898--2910. Association for Computational
  Linguistics.

\bibitem[{Li et~al.(2023{\natexlab{a}})Li, Liu, Xiong, Yu, Gu, Liu, and
  Yu}]{SANTA}
Xinze Li, Zhenghao Liu, Chenyan Xiong, Shi Yu, Yu~Gu, Zhiyuan Liu, and Ge~Yu.
  2023{\natexlab{a}}.
\newblock \href {https://doi.org/10.18653/v1/2023.findings-acl.734}
  {Structure-aware language model pretraining improves dense retrieval on
  structured data}.
\newblock In \emph{{ACL} (Findings)}, pages 11560--11574. Association for
  Computational Linguistics.

\bibitem[{Li et~al.(2025)Li, Wang, Liu, Yu, Wang, Yan, Fu, Gu, and Yu}]{CONAN}
Xinze Li, Hanbin Wang, Zhenghao Liu, Shi Yu, Shuo Wang, Yukun Yan, Yukai Fu,
  Yu~Gu, and Ge~Yu. 2025.
\newblock \href {https://doi.org/10.1145/3695868} {Building a coding assistant
  via the retrieval-augmented language model}.
\newblock \emph{{ACM} Trans. Inf. Syst.}, 43(2):39:1--39:25.

\bibitem[{Li et~al.(2021{\natexlab{b}})Li, Liu, Xiong, and Liu}]{DANCE}
Yizhi Li, Zhenghao Liu, Chenyan Xiong, and Zhiyuan Liu. 2021{\natexlab{b}}.
\newblock \href {https://doi.org/10.1145/3471158.3472245} {More robust dense
  retrieval with contrastive dual learning}.
\newblock In \emph{{ICTIR}}, pages 287--296. {ACM}.

\bibitem[{Li et~al.(2023{\natexlab{b}})Li, Zhang, Zhang, Long, Xie, and
  Zhang}]{gte-Qwen2-Instruct}
Zehan Li, Xin Zhang, Yanzhao Zhang, Dingkun Long, Pengjun Xie, and Meishan
  Zhang. 2023{\natexlab{b}}.
\newblock \href {https://doi.org/10.48550/arXiv.2308.03281} {Towards general
  text embeddings with multi-stage contrastive learning}.
\newblock \emph{CoRR}, abs/2308.03281.

\bibitem[{Nguyen et~al.(2016)Nguyen, Rosenberg, Song, Gao, Tiwary, Majumder,
  and Deng}]{MSMARCO}
Tri Nguyen, Mir Rosenberg, Xia Song, Jianfeng Gao, Saurabh Tiwary, Rangan
  Majumder, and Li~Deng. 2016.
\newblock \href {https://ceur-ws.org/Vol-1773/CoCoNIPS\_2016\_paper9.pdf} {{MS}
  {MARCO:} {A} human generated machine reading comprehension dataset}.
\newblock In \emph{CoCo@NIPS}, volume 1773 of \emph{{CEUR} Workshop
  Proceedings}. CEUR-WS.org.

\bibitem[{Rao et~al.(2023)Rao, Ding, Qi, Fang, Liu, Shen, and
  Tao}]{rao2023dynamic}
Jun Rao, Liang Ding, Shuhan Qi, Meng Fang, Yang Liu, Li~Shen, and Dacheng Tao.
  2023.
\newblock \href {https://ieeexplore.ieee.org/document/10102558} {Dynamic
  contrastive distillation for image-text retrieval}.
\newblock \emph{IEEE Transactions on Multimedia}, 25:8383--8395.

\bibitem[{Rao et~al.(2025)Rao, Lin, Liu, Ke, Lian, Jin, Cheng, Yu, and
  Zhang}]{apt}
Jun Rao, Zepeng Lin, Xuebo Liu, Xiaopeng Ke, Lian Lian, Dong Jin, Shengjun
  Cheng, Jun Yu, and Min Zhang. 2025.
\newblock \href {https://aclanthology.org/2025.findings-acl.1079/} {{APT:}
  improving specialist {LLM} performance with weakness case acquisition and
  iterative preference training}.
\newblock In \emph{{ACL} (Findings)}, pages 20958--20980. Association for
  Computational Linguistics.

\bibitem[{Rao et~al.(2022)Rao, Wang, Ding, Qi, Zhan, Liu, and
  Tao}]{rao2022reproducibility}
Jun Rao, Fei Wang, Liang Ding, Shuhan Qi, Yibing Zhan, Weifeng Liu, and Dacheng
  Tao. 2022.
\newblock \href {https://arxiv.org/pdf/2203.03853} {Where does the performance
  improvement come from - a reproducibility concern about image-text
  retrieval}.
\newblock In \emph{SIGIR}.

\bibitem[{Reddy et~al.(2022)Reddy, M{\`{a}}rquez, Valero, Rao, Zaragoza,
  Bandyopadhyay, Biswas, Xing, and Subbian}]{ProductSearch}
Chandan~K. Reddy, Llu{\'{\i}}s M{\`{a}}rquez, Fran Valero, Nikhil Rao, Hugo
  Zaragoza, Sambaran Bandyopadhyay, Arnab Biswas, Anlu Xing, and Karthik
  Subbian. 2022.
\newblock \href {https://doi.org/10.48550/arXiv.2206.06588} {Shopping queries
  dataset: {A} large-scale {ESCI} benchmark for improving product search}.
\newblock \emph{CoRR}, abs/2206.06588.

\bibitem[{Sciavolino et~al.(2021)Sciavolino, Zhong, Lee, and Chen}]{mask_entiy}
Christopher Sciavolino, Zexuan Zhong, Jinhyuk Lee, and Danqi Chen. 2021.
\newblock \href {https://doi.org/10.18653/v1/2021.emnlp-main.496} {Simple
  entity-centric questions challenge dense retrievers}.
\newblock In \emph{{EMNLP} {(1)}}, pages 6138--6148. Association for
  Computational Linguistics.

\bibitem[{Sun et~al.(2025)Sun, Guo, Liu, Zhang, Hou, and Li}]{sun2025zero}
Dong Sun, Wenya Guo, Xumeng Liu, Ying Zhang, Zhaoxiang Hou, and Zengxiang Li.
  2025.
\newblock Zero-shot document retrieval with hybrid pseudo-document retriever.
\newblock In \emph{ICASSP 2025-2025 IEEE International Conference on Acoustics,
  Speech and Signal Processing (ICASSP)}, pages 1--5. IEEE.

\bibitem[{Tan et~al.(2025)Tan, Dou, Wang, Wang, Chen, and Wen}]{html}
Jiejun Tan, Zhicheng Dou, Wen Wang, Mang Wang, Weipeng Chen, and Ji{-}Rong Wen.
  2025.
\newblock \href {https://doi.org/10.1145/3696410.3714546} {Htmlrag: {HTML} is
  better than plain text for modeling retrieved knowledge in {RAG} systems}.
\newblock In \emph{{WWW}}, pages 1733--1746. {ACM}.

\bibitem[{Thakur et~al.(2021)Thakur, Reimers, R{\"{u}}ckl{\'{e}}, Srivastava,
  and Gurevych}]{BEIR}
Nandan Thakur, Nils Reimers, Andreas R{\"{u}}ckl{\'{e}}, Abhishek Srivastava,
  and Iryna Gurevych. 2021.
\newblock \href
  {https://datasets-benchmarks-proceedings.neurips.cc/paper/2021/hash/65b9eea6e1cc6bb9f0cd2a47751a186f-Abstract-round2.html}
  {{BEIR:} {A} heterogeneous benchmark for zero-shot evaluation of information
  retrieval models}.
\newblock In \emph{NeurIPS Datasets and Benchmarks}.

\bibitem[{Wang et~al.(2024)Wang, Yang, Huang, Yang, Majumder, and
  Wei}]{wang2024multilingual}
Liang Wang, Nan Yang, Xiaolong Huang, Linjun Yang, Rangan Majumder, and Furu
  Wei. 2024.
\newblock \href {https://doi.org/10.48550/arXiv.2402.05672} {Multilingual {E5}
  text embeddings: {A} technical report}.
\newblock \emph{CoRR}, abs/2402.05672.

\bibitem[{Wang et~al.(2021)Wang, Wang, Joty, and Hoi}]{CodeT5}
Yue Wang, Weishi Wang, Shafiq~R. Joty, and Steven C.~H. Hoi. 2021.
\newblock \href {https://doi.org/10.18653/v1/2021.emnlp-main.685} {Codet5:
  Identifier-aware unified pre-trained encoder-decoder models for code
  understanding and generation}.
\newblock In \emph{{EMNLP} {(1)}}, pages 8696--8708. Association for
  Computational Linguistics.

\bibitem[{Warner et~al.(2024)Warner, Chaffin, Clavié, Weller, Hallström,
  Taghadouini, Gallagher, Biswas, Ladhak, Aarsen, Cooper, Adams, Howard, and
  Poli}]{modernBERT}
Benjamin Warner, Antoine Chaffin, Benjamin Clavié, Orion Weller, Oskar
  Hallström, Said Taghadouini, Alexis Gallagher, Raja Biswas, Faisal Ladhak,
  Tom Aarsen, Nathan Cooper, Griffin Adams, Jeremy Howard, and Iacopo Poli.
  2024.
\newblock \href {https://arxiv.org/abs/2412.13663} {Smarter, better, faster,
  longer: A modern bidirectional encoder for fast, memory efficient, and long
  context finetuning and inference}.
\newblock \emph{CoRR}, abs/2412.13663.

\bibitem[{Xiao et~al.(2023)Xiao, Liu, Zhang, and Muennighoff}]{bge_embedding}
Shitao Xiao, Zheng Liu, Peitian Zhang, and Niklas Muennighoff. 2023.
\newblock \href {https://doi.org/10.48550/arXiv.2309.07597} {C-pack: Packaged
  resources to advance general chinese embedding}.
\newblock \emph{CoRR}, abs/2309.07597.

\bibitem[{Xiong et~al.(2021{\natexlab{a}})Xiong, Xiong, Li, Tang, Liu, Bennett,
  Ahmed, and Overwijk}]{ANCE}
Lee Xiong, Chenyan Xiong, Ye~Li, Kwok{-}Fung Tang, Jialin Liu, Paul~N. Bennett,
  Junaid Ahmed, and Arnold Overwijk. 2021{\natexlab{a}}.
\newblock \href {https://openreview.net/forum?id=zeFrfgyZln} {Approximate
  nearest neighbor negative contrastive learning for dense text retrieval}.
\newblock In \emph{{ICLR}}. OpenReview.net.

\bibitem[{Xiong et~al.(2021{\natexlab{b}})Xiong, Li, Iyer, Du, Lewis, Wang,
  Mehdad, Yih, Riedel, Kiela, and Oguz}]{MDR}
Wenhan Xiong, Xiang~Lorraine Li, Srini Iyer, Jingfei Du, Patrick S.~H. Lewis,
  William~Yang Wang, Yashar Mehdad, Scott Yih, Sebastian Riedel, Douwe Kiela,
  and Barlas Oguz. 2021{\natexlab{b}}.
\newblock \href {https://openreview.net/forum?id=EMHoBG0avc1} {Answering
  complex open-domain questions with multi-hop dense retrieval}.
\newblock In \emph{{ICLR}}. OpenReview.net.

\bibitem[{Yang et~al.(2023)Yang, Liu, Li, Sun, and Xie}]{Longtriever}
Junhan Yang, Zheng Liu, Chaozhuo Li, Guangzhong Sun, and Xing Xie. 2023.
\newblock \href {https://doi.org/10.18653/v1/2023.emnlp-main.223} {Longtriever:
  a pre-trained long text encoder for dense document retrieval}.
\newblock In \emph{{EMNLP}}, pages 3655--3665. Association for Computational
  Linguistics.

\bibitem[{Zhao et~al.(2024)Zhao, Liu, Ren, and Wen}]{survey}
Wayne~Xin Zhao, Jing Liu, Ruiyang Ren, and Ji{-}Rong Wen. 2024.
\newblock \href {https://doi.org/10.1145/3637870} {Dense text retrieval based
  on pretrained language models: {A} survey}.
\newblock \emph{{ACM} Trans. Inf. Syst.}, 42(4):89:1--89:60.

\end{thebibliography}

\end{document}